\documentclass[11pt,letterpaper]{article}

\usepackage{times}
\usepackage{relsize}
\usepackage{microtype}
\usepackage{cite}
\usepackage{xspace}
\usepackage[table,xcdraw]{xcolor}
\usepackage{color,colortbl}
\usepackage{subfigure}
\usepackage{alltt}
\usepackage{url}
\usepackage{comment}
\usepackage{graphicx}
\usepackage{relsize}
\usepackage{paralist}
\usepackage{todonotes}
\usepackage{wrapfig}
\usepackage{titlesec}
\usepackage{epsfig}
\usepackage{geometry}
\usepackage{tikz}
\usepackage[normalem]{ulem}
\usepackage{booktabs}
\usepackage{wrapfig}
\usepackage{multirow}
\usepackage{pifont}
\usepackage{wrapfig}
\usepackage{soul}
\usepackage{enumitem}
\usepackage{tcolorbox}
\usepackage{pdfpages}
\usepackage{fancyhdr}
\usepackage{longtable}
\usepackage{makecell}
\usepackage{lettrine}
\usepackage{listings}
\usepackage{longtable}

\titlespacing{\section}{0pc}{2pc}{1pc}
\titlespacing{\subsection}{0pc}{2pc}{1pc}
\titlespacing{\subsubsection}{0pc}{1pc}{1pc}
\titleformat*{\section}{\Large\bfseries}
\titleformat*{\subsection}{\large\bfseries}
\titleformat*{\subsubsection}{\normalsize\bfseries}

\lstdefinestyle{wcsStyle}{
  tabsize=2,
  showspaces=false,
  showstringspaces=false,
  aboveskip=0em,
  belowskip=0em,
}

\definecolor{Gray}{gray}{0.9}

\newcommand{\pp}[1]{\medskip \noindent \textbf{\emph{#1.}}\xspace}

\newcommand{\bp}[2]{\begin{tabular}{@{\textbullet~}p{#1}@{}}#2\end{tabular}}

\newif\ifdraft
\drafttrue

\ifdraft{}
  \newcommand{\jhanote}[1]{ {\textcolor{red} { ***shantenu: #1 }}}
\else
  \newcommand{\jhanote}[1]{}
\fi

\begin{document}

\includepdf[pages=-]{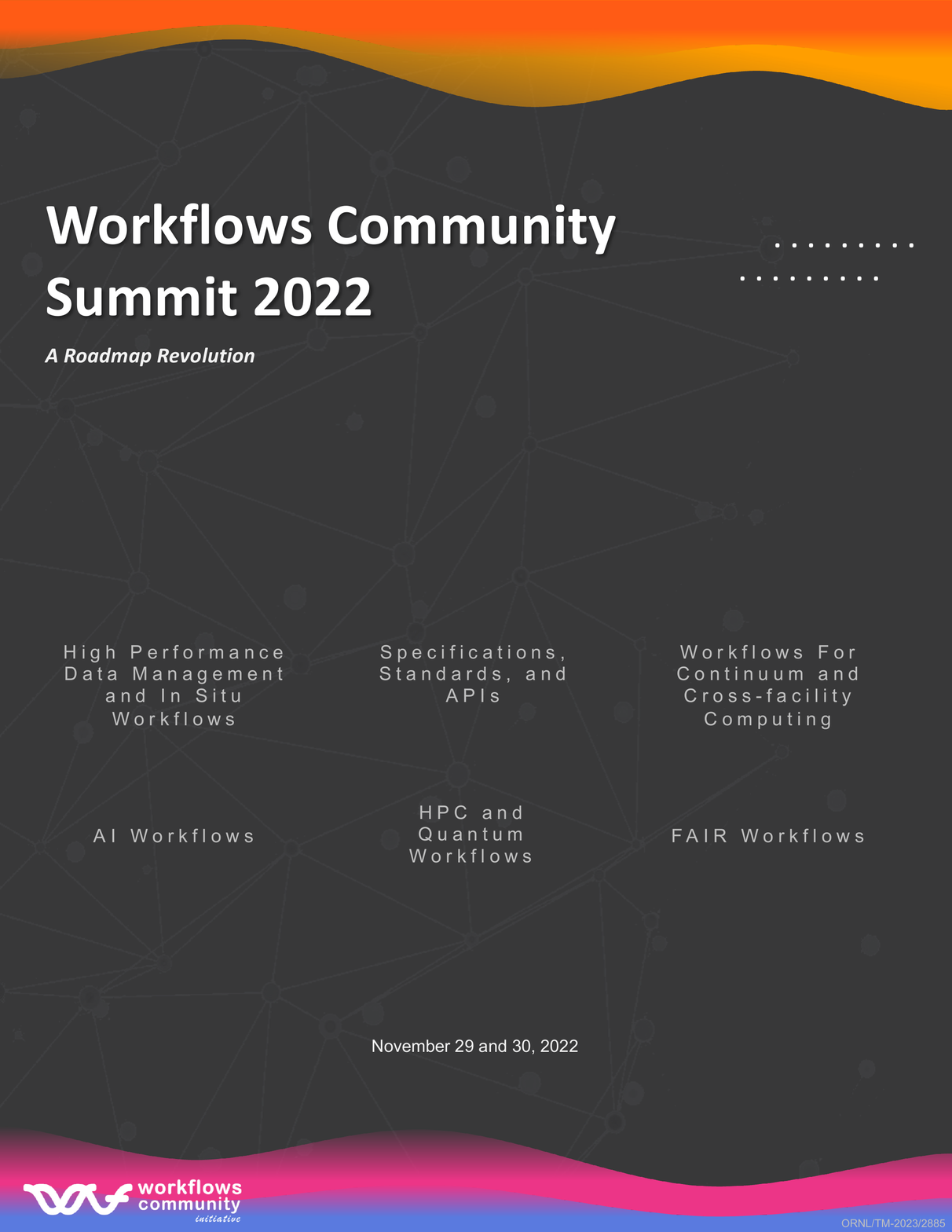}


\pagestyle{fancy}
\fancyhf{}
\rhead{
  \includegraphics[height=11pt]{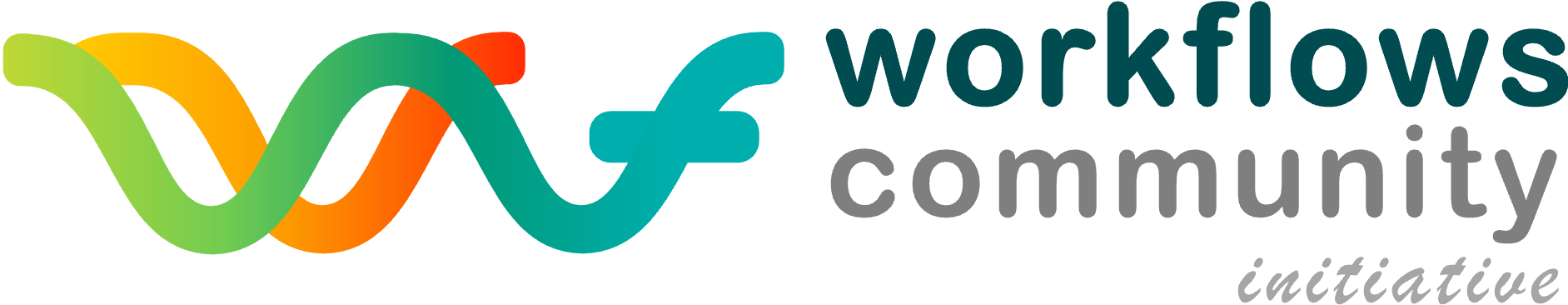}
}
\lhead{Workflows Community Summit 2022}
\rfoot{\thepage}


\begin{table}[!ht]
\centering
\small
\begin{tabular}{p{16cm}}
    \vspace{2em}
    \textbf{Disclaimer}
    \\
    The Workflows Community Summit was supported by the Exascale Computing Project (17-SC-20-SC), a collaborative effort of the U.S. Department of Energy Office of Science and the National Nuclear Security Administration. This research used resources of the Oak Ridge Leadership Computing Facility at the Oak Ridge National Laboratory, which is supported by the Office of Science of the U.S. Department of Energy under Contract No. DE-AC05-00OR22725.
    \\ \vspace{0.5em}
    This report was prepared as an account of work sponsored by agencies of the United States Government. Neither the United States Government nor any agency thereof, nor any of their employees, makes any warranty, express or implied, or assumes any legal liability or responsibility for the accuracy, completeness, or usefulness of any information, apparatus, product, or process disclosed, or represents that its use would not infringe privately owned rights. Reference herein to any specific commercial product, process, or service by trade name, trademark, manufacturer, or otherwise, does not necessarily constitute or imply its endorsement, recommendation, or favoring by the United States Government or any agency thereof. The views and opinions of authors expressed herein do not necessarily state or reflect those of the United States Government or any agency thereof.
    \\
    \vspace{0.5em}
    \textbf{License}
    \\
    This report is made available under a Creative Commons Attribution 4.0 International Public license ({\small \url{https://creativecommons.org/licenses/by/4.0}}).
\end{tabular}
\end{table}
\vspace*{\fill}

\newpage

\begin{table}[!ht]
\centering
\smaller
\begin{tabular}{p{16cm}}
    \vspace{2em}
    \textbf{\normalsize Preferred citation} 
    \\
    R. Ferreira da Silva, R.M. Badia, V. Bala, D. Bard, T. Bremer, I. Buckley, S. Caino-Lores, K. Chard, C. Goble, S. Jha
% ...
    ``\emph{Workflows Community Summit 2022: A Roadmap Revolution}", Technical Report, ORNL/TM-2023/2885, March 2023, DOI: 10.5281/zenodo.7750670.
    \\
    \rowcolor[HTML]{F7F7F7}
    \lstset{basicstyle=\scriptsize,style=wcsStyle}
    \begin{lstlisting}
@techreport{wcs2022,
  author = {Ferreira da Silva, Rafael and Badia, Rosa M. and Bala, Venkat and Bard, Debbie 
            and Bremer, Timo and Buckley, Ian and Caino-Lores, Silvina and Chard, Kyle and 
            Goble, Carole and Jha, Shantenu and Katz, Daniel S. and Laney, Daniel and 
            Parashar, Manish and Suter, Frederic and Tyler, Nick and Uram, Thomas and 
            Altintas, Ilkay and Andersson, Stefan and Arndt, William and Aznar, Juan and 
            Bader, Jonathan and Balis, Bartosz and Blanton, Chris and Braghetto, Kelly Rosa 
            and Brodutch, Aharon  and Brunk, Paul and Casanova, Henri and Cervera Lierta, Alba 
            and Chigu, Justin and Coleman, Taina and Collier, Nick and Colonnelli, Iacopo and 
            Coppens, Frederik and Crusoe, Michael and Cunningham, Will and de Paula Kinoshita, 
            Bruno and Di Tommaso, Paolo and Doutriaux, Charles and Downton, Matthew and Elwasif, 
            Wael and Enders, Bjoern and Erdmann, Chris and Fahringer, Thomas and Figueiredo, 
            Ludmilla and Filgueira, Rosa and Foltin, Martin and Fouilloux, Anne and Gadelha, 
            Luiz and Gallo, Andy and Garcia, Artur and Garijo, Daniel and Gerlach, Roman and 
            Grant, Ryan and Grayson, Samuel and Grubel, Patricia and Gustafsson, Johan and 
            Hayot, Valerie and Hernandez, Oscar and Hilbrich, Marcus and Justine, Annmary and 
            Laflotte, Ian, and Lehmann, Fabian and Luckow, Andre and Luettgau, Jakob and 
            Maheshwari, Ketan and Matsuda, Motohiko and Medic, Doriana and Mendygral, Pete and 
            Michalewicz, Marek and Nonaka, Jorji and Pawlik, Maciej and Pottier, Loic and 
            Pouchard, Line and Putz, Mathias and Radha, Santosh Kumar and Ramakrishnan, Lavanya 
            and Ristov, Sashko and Romano, Paul and Rosendo, Daniel and Ruefenacht, Martin and 
            Rycerz, Katarzyna and Saurabh, Nishant and Savchenko, Volodymyr and Schulz, Martin 
            and Simpson, Christine and Sirvent, Raul and Skluzacek, Tyler and Soiland-Reyes, 
            Stian and Souza, Renan and Sukumar, Sreenivas Rangan and Sun, Ziheng and Sussman, 
            Alan and Thain, Douglas and Titov, Mikhail and Tovar, Benjamin and Tripathy, Aalap 
            and Turilli, Matteo and Tuznik, Bartosz and van Dam, Hubertus and Vivas, Aurelio 
            and Ward, Logan and Widener, Patrick and Wilkinson, Sean R. and Zawalska, Justyna 
            and Zulfiqar, Mahnoor},
  title       = {{Workflows Community Summit 2022: A Roadmap Revolution}},
  month       = {March},
  year        = {2023},
  publisher   = {Zenodo},
  number      = {ORNL/TM-2023/2885},
  doi         = {10.5281/zenodo.7750670},
  institution = {Oak Ridge National Laboratory}
}
    \end{lstlisting}
    \\
\end{tabular}
\end{table}
\vspace*{\fill}

\newpage


\newpage


\section{Introduction}
\label{sec:introduction}

Scientific workflows have become integral tools in broad scientific computing use cases~\cite{ben2020workflows}. Science discovery is increasingly dependent on workflows to orchestrate large and complex scientific experiments that range from the execution of a cloud-based data preprocessing pipeline to multi-facility instrument-to-edge-to-HPC computational workflows~\cite{badia2017workflows, ferreiradasilva-fgcs-2017}. Given the changing landscape of scientific computing (often referred to as a computing continuum~\cite{foster19continuum}) and the evolving needs of emerging scientific applications, it is paramount that the development of novel scientific workflows and system functionalities seek to increase the efficiency, resilience, and pervasiveness of existing systems and applications. Specifically, the proliferation of machine learning/artificial intelligence (ML/AI) workflows, need for processing large-scale datasets produced by instruments at the edge, intensification of near real-time data processing, support for long-term experiment campaigns, and emergence of quantum computing as an adjunct to HPC, have significantly changed the functional and operational requirements of workflow systems. Workflow systems now need to, for example, support data streams from the edge-to-cloud-to-HPC~\cite{altintas2022towards}, enable the management of many small-sized files~\cite{ferreiradasilva2021works}, allow data reduction while ensuring high accuracy~\cite{poeschel2022transitioning}, orchestrate distributed services (workflows, instruments, data movement, provenance, publication, etc.) across computing and user facilities~\cite{antypas2021enabling}, among others. Further, to accelerate science, it is also necessary that these systems implement specifications/standards and APIs for seamless (horizontal and vertical) integration between systems and applications~\cite{cohen2017scientific, al2021exaworks}, as well as enable the publication of workflows and their associated products according to the FAIR principles~\cite{goble2020fair}.

\subsection{Summit Organization}

\begin{figure}[!t]
    \centering
    \includegraphics[width=.88\linewidth]{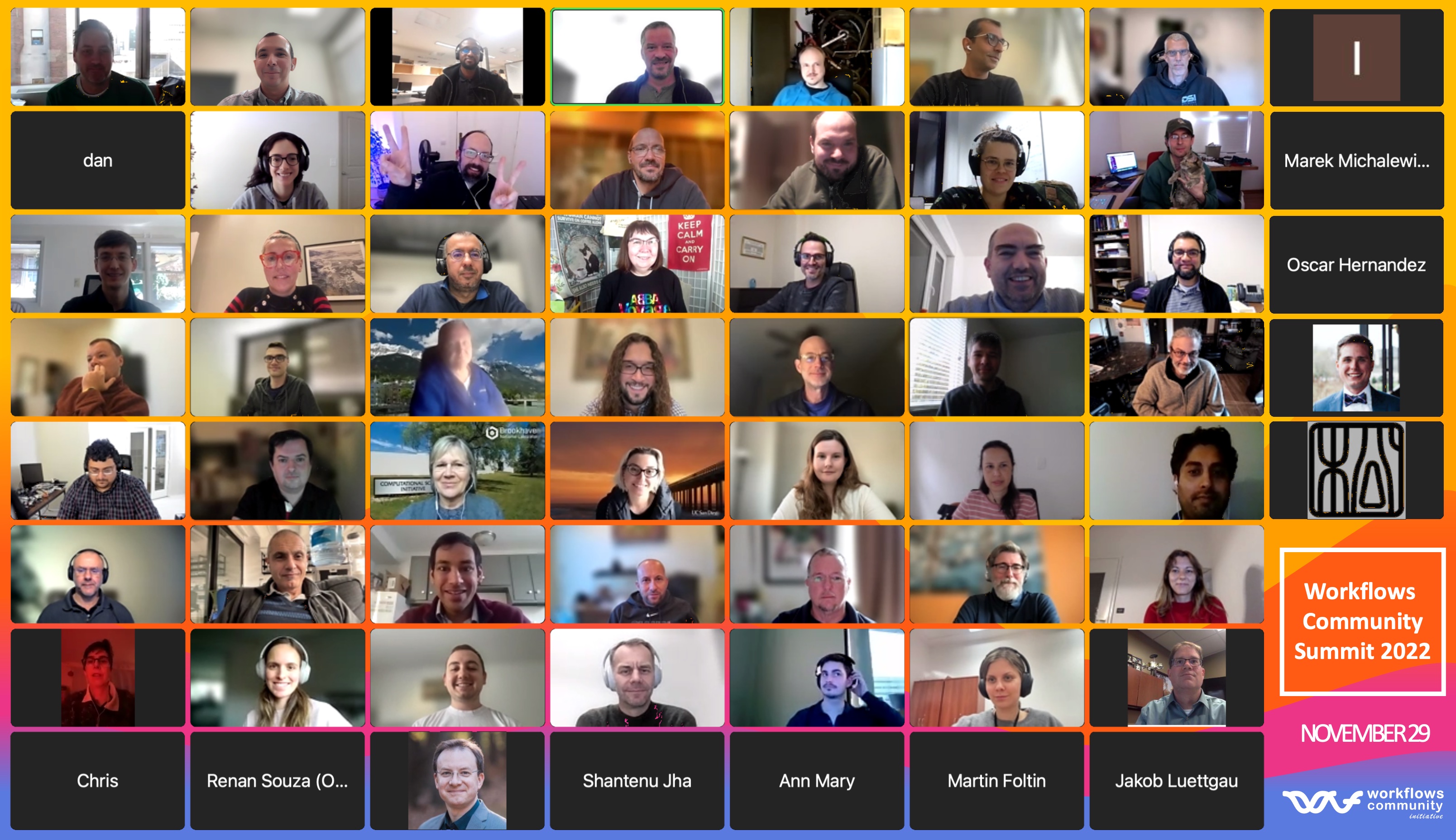} \\
    \vspace{5pt}
    \includegraphics[width=.88\linewidth]{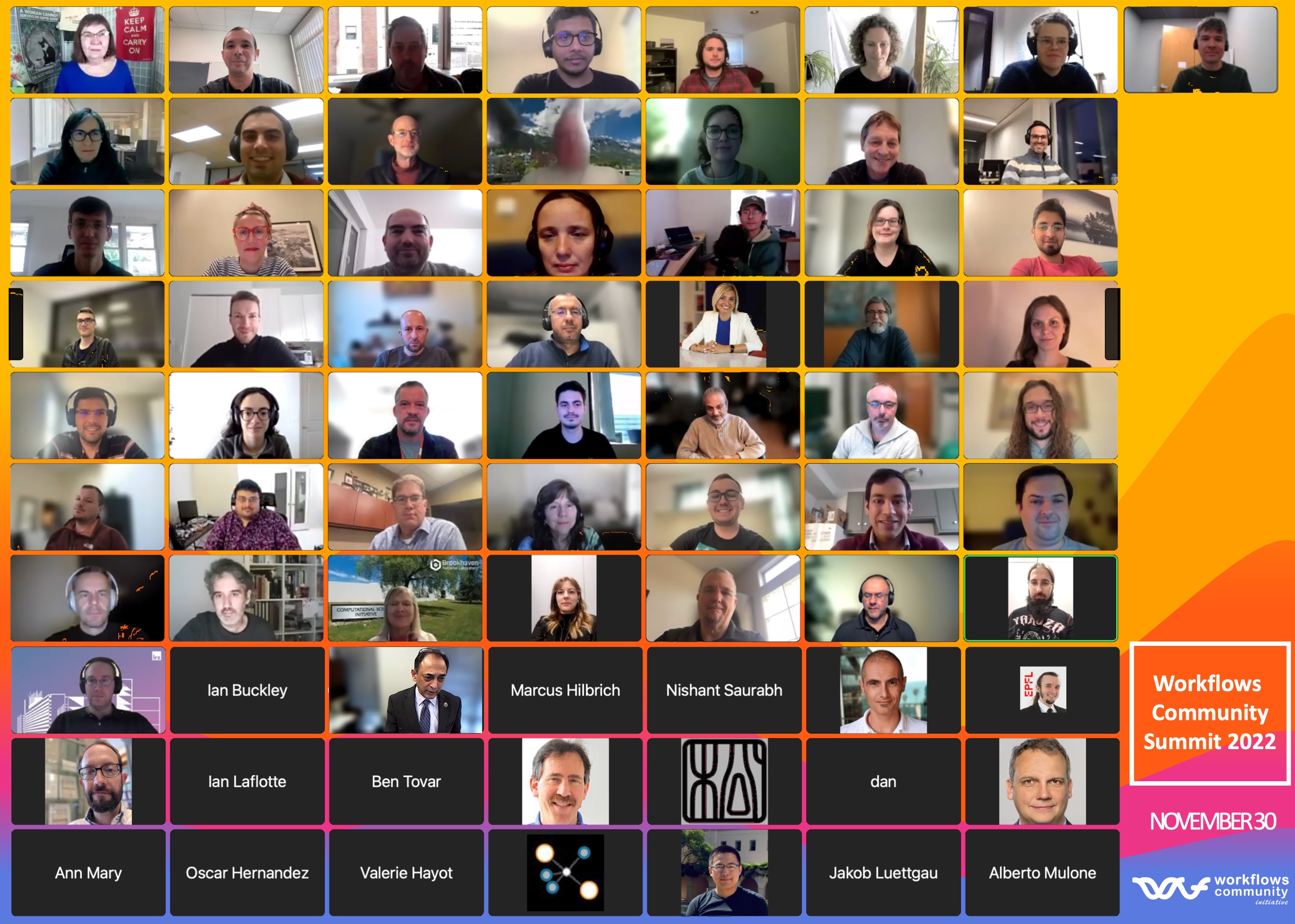}
    \caption{Screenshot of the 2022 edition of the Workflows Community Summit participants. (The event was held virtually via Zoom on November 29 and 30, 2022.)}
    \label{fig:participants}
\end{figure}

This document reports on discussions and findings from the 2022 international edition of the ``Workflows Community Summit" that took place on November 29 and 30, 2022~\cite{wcs-2022}.  The two-day summit included 106 participants (Figure~\ref{fig:participants}), from a group of international researchers and developers (Australia, Austria, Belgium, Brazil, Canada, Colombia, Egypt, France, Germany, Italy, Japan, Netherlands, Norway, Poland, Spain, Switzerland, United Kingdom, United States) from distinct workflow management systems and users, and representatives from funding agencies and industry.

The summit was organized by lead members of the Workflows Community Initiative (WCI)~\cite{wci}, a volunteer effort that brings the workflows community together (users, developers, researchers, and facilities) to provide community resources and capabilities to enable scientists and workflow systems developers to discover software products, related efforts, events, technical reports, etc. and engage in community-wide efforts to tackle workflows grand challenges. These members efforts have been supported by distinct research projects that contributed to the organization of the event, including ExaWorks~\cite{al2021exaworks}, WorkflowHub~\cite{goble2021implementing}, eFlows4HPC~\cite{eflows4hpc}, and Covalent~\cite{covalent}.

\subsection{Summit Structure and Activities}

Based on the outcomes of the 2021 summits~\cite{wcs2021, wcs2021technical, wcs2021facilities}, the community developed a roadmap for workflows research and development~\cite{ferreiradasilva2021works} (summarized in Section~\ref{sec:roadmap}). From this roadmap, we have identified six cross-cutting research/technical topics for the 2022 edition of the summit, each of which was the object of a focused discussion led by a volunteer community member (Table~\ref{tab:topics}). (Note that the first three topics were discussed in the first day of the event, and the latter three topics in the second day.) For each of the technical topics shown in Table~\ref{tab:topics}, the co-leaders gave a plenary 10-minute lightning talk followed by focused discussions in breakout sessions. The goal of these sessions was to (i)~review the previous challenges and proposed solutions; and (ii)~identify crucial gaps and potential short- and long-term solutions for enabling emerging and new workflow applications given the rapid evolution of the computing continuum paradigm. The lead then reported on the outcome of the discussion in plenary sessions, after which final remarks were given by the organizers and the summit was adjourned. Additionally, the summit included an invited talk from \textbf{\emph{Prof. Manish Parashar}} (University of Utah), a renowned researcher in the field of Scientific Computing, entitled ``A Translational Perspective on End-to-End Workflows." The cross-cutting theme of Parashar's talk was discussed throughout all breakout sessions.

\begin{table}[!t]
    \centering
    \small
    \setlength{\tabcolsep}{12pt}
    \begin{tabular}{p{6.5cm}p{8cm}}
        \toprule
        \textbf{Topic} & \textbf{Discussion Co-Leaders} \\
        \midrule
        \rowcolor[HTML]{F2F2F2}
        Topic 1: Specifications, standards, and APIs & Daniel Laney (LLNL), Kyle Chard (UChicago/ANL) \\
        Topic 2: AI workflows & Shantenu Jha (BNL), Timo Bremer (LLNL) \\
        \rowcolor[HTML]{F2F2F2}
        Topic 3: High performance data management and in situ workflows & Silvina Caino-Lores (UTK), Fr\'ed\'eric Suter (ORNL) \\
        Topic 4: HPC and Quantum workflows & Rosa M. Badia (BSC), Ian Buckley (Agnostiq), Venkat Bala (Agnostiq) \\
        \rowcolor[HTML]{F2F2F2}
        Topic 5: FAIR computational workflows & Carole Goble (UManchester), Daniel S. Katz (UIUC) \\
        Topic 6: Workflows for continuum and cross-facility computing & Debbie Bard (NERSC), Tom Uram (ANL), Nick Tyler (NERSC) \\
        \bottomrule
    \end{tabular}
    \caption{Workflows Community Summit topics and discussion co-leaders.}
    \label{tab:topics}
\end{table}

\medskip

All presentations and videos can be found in the summit website (\url{https://workflows.community/summits/2022}), and videos can be watched from WCI's YouTube channel (\url{https://www.youtube.com/playlist?list=PLAtmuqHExRvO-l-mACmlK_4mfmG6yLnpN}).

\newpage
\section{An Overview of the 2021 Community Roadmap for Scientific Workflows Research and Development}
\label{sec:roadmap}

\begin{table*}[!b]
\centering
\scriptsize
\begin{tabular}{p{1.3cm}p{6.9cm}p{6.9cm}}
\toprule 
Theme & Challenges & Community Activities \\
\midrule
\makecell[l]{FAIR\\Computational\\Workflows} &
\bp{6.9cm}{
    FAIR principles for computational workflows that consider the complex lifecycle from specification to execution and data products
    \\
    Metrics to measure the ``FAIRness'' of a workflow
    \\
    Principles, policies, and best practices
}
&
\bp{6.9cm}{
    Review prior and current efforts for FAIR data and software with respect to workflows, and outline principles for FAIR workflows
    \\
    Define recommendations for FAIR workflow developers and systems
    \\
    Automate FAIRness in workflows by recording necessary provenance data
}
\\
\midrule

\makecell[l]{AI\\Workflows} &
\bp{6.9cm}{
    Support for heterogeneous compute resources and fine-grained data management features, versioning, and data provenance capabilities
    \\
    Capabilities for enabling workflow steering and dynamic workflows
    \\
    Integration of ML frameworks into the current HPC landscape
}
&
\bp{6.9cm}{
    Develop comprehensive use cases for sample problems with representative workflow structures and data types
    \\
    Define a process for characterizing the challenges for enabling AI workflows
    \\
    Develop AI workflows as a way to benchmark HPC systems
}
\\
\midrule

\makecell[l]{Exascale\\Challenges\\and Beyond} &
\bp{6.9cm}{
    Resource allocation policies and schedulers are not designed for workflow-aware abstractions, thus users tend to use an ill-fitted job abstraction
    \\
    Unfavorable design of resource descriptions and mechanisms for workflow users/systems, and lack of fault-tolerance and fault-recovery solutions
}
&
\bp{6.9cm}{
    Develop documentation in the form of workflow templates/recipes/miniapps for execution on high-end HPC systems
    \\
    Specify benchmark workflows for exascale execution
    \\
    Include workflow requirements as part of the machine procurement process
}
\\
\midrule

\makecell[l]{APIs, Reuse,\\Interoperability,\\and Standards} &
\bp{6.9cm}{
    Workflow systems differ by design, thus interoperability at some layers is likely to be more impactful than others
    \\
    Workflow standards are typically developed by a subset of the community
    \\
    Quantifying the value of common representations of workflows is not trivial
}
&
\bp{6.9cm}{
    Identify differences and commonalities between different systems
    \\
    Identify and characterize domain-specific efforts, identify workflow patterns, and develop case-studies of business process workflows and serverless workflow systems
}
\\
\midrule

\makecell[l]{Training\\and Education} &
\bp{6.9cm}{
    Many workflow systems have high barriers to entry and lack training materials
    \\
    Homegrown workflow solutions and constraints can prevent users from reproducing their functionality on workflow systems developed by others
    \\
    Unawareness of the workflow technological and conceptual landscape
}
&
\bp{6.9cm}{
    Identify basic sample workflow patterns, develop a community workflow knowledge-base, and look at current research on technology adoption
    \\
    Include workflow terminology and concepts in university curricula and software carpentry efforts
}
\\
\midrule

\makecell[l]{Building\\a Workflows\\Community} &
\bp{6.9cm}{
    Diverse definitions of a ``workflows community''
    \\
    Remedy the inability to link developers and users to bridge translational gaps
    \\
    Pathways for participation in a network of researchers, developers, and users
}
& 
\bp{6.9cm}{
    Establish a common knowledge-base for workflow technology
    \\
    Establish a \emph{Workflow Guild}: an organization focused on interaction and relationships, providing self-support between workflow developers and their systems
}
\\
\bottomrule
\end{tabular}
\caption{Summary of workflows research and development challenges and proposed community activities identified during the 2021 summits~\cite{ferreiradasilva2021works}.}
\label{tab:challenges}
\end{table*}

The overarching goal of the 2021 Workflows Community Summits was to (i)~develop a view of the state of the art, (ii)~identify key research challenges, (iii)~articulate a vision for potential activities, and (iv)~explore technical approaches for realizing (part of) this vision. The summits gathered lead researchers and developers from around the world, and spanning distinct workflow systems and user communities. We synthesized the discussions and outcomes of the previous summits and findings into a community roadmap~\cite{ferreiradasilva2021works} that presents a consolidated view of the state of the art, challenges, and  potential activities. Table~\ref{tab:challenges} presents, in the form of top-level themes, a summary of those challenges and targeted community activities. Table~\ref{tab:roadmap} summarizes a proposed community roadmap with technical approaches. 

\begin{table*}[!t]
\centering
\scriptsize
\begin{tabular}{p{3.2cm}p{12.3cm}}
\toprule 
Thrust & Roadmap Milestones \\
\midrule
\makecell[l]{Definition of common\\workflow patterns and\\benchmarks} &
\bp{12.3cm}{
    Define small sets of workflow patterns and benchmark deliverables, and implement them using a selected set of workflow systems
    \\
    Investigate automatic generation of patterns and configurable benchmarks (e.g., to enable weak and strong scaling experiments)
    \\
    Establish or leverage a centralized repository to host and curate patterns and benchmarks
}
\\
\midrule

\makecell[l]{Identifying paths toward\\interoperability of workflow\\systems} &
\bp{12.3cm}{
    Define interoperability for different roles, develop a horizontal interoperability (i.e., making interoperable components), and establish a requirements document per abstraction layer
    \\
    Develop real-world workflow benchmarks, use cases for interoperability, and common APIs that represent workflow library components
    \\
    Establish a workflow systems developer community
}
\\
\midrule

\makecell[l]{Improving workflow systems'\\interface with legacy and\\ emerging HPC software and\\hardware stacks} &
\bp{12.3cm}{
    Document a machine-readable description of key properties of widely used sites, and remote authentication needs from the workflow perspective
    \\
    Identify new workflow patterns (e.g., motivated by AI workflows), attain portability across heterogeneous hardware, and develop a registry of execution environment information
    \\
    Organize a community event involving workflow system developers, end users, authentication technology providers, and facility operators
}
\\
\bottomrule
\end{tabular}
\caption{Summary of technical roadmap milestones per research and development thrust proposed during the 2021 summits~\cite{ferreiradasilva2021works}.}
\label{tab:roadmap}
\end{table*}

\medskip

In the 2022 edition of the Workflows Community Summit, we revisited this roadmap, discussed the relevance of the challenges and proposed milestones given the current landscape of workflow applications needs and emerging infrastructures. In the following sections, we summarize the outcomes of these discussions.


\section{Specifications, Standards, and APIs}
\label{sec:specifications}

Defining reference specifications, standards, and APIs is paramount to enable interoperability across the workflow software stack; for capturing end-to-end workflow provenance to facilitate user understanding of the scientific output; for enabling composability of underlying workflow system components to enable seamless exchange of functional modules; among others.
The workflows community has tackled this problem in many ways ranging from solutions that provide reference implementations of standards and/or specifications~\cite{crusoe2022methods, al2021exaworks} to define common patterns/motifs~\cite{gallo2023industrial, garijo2014common}. In this breakout session, participants focused on identifying the challenges impeding adoption of current standards, specifications, best practices, and APIs developed by the community. Below, we summarize the key outcomes of these discussions, highlighting exemplar efforts, and outline proposed actionable solutions to be tackled by the community.

\subsection{Workflow Stack}
\label{sec:workflow-stack}

Despite the vast efforts for providing reference standard implementations and identifying common patterns, defining the levels of the workflow stack which are amenable to sharing technologies (and potentially APIs) is still an open question and potentially one of the critical aspects hindering wide adoption of proposed standards and specifications. For instance, there are several ways to define a workflow stack: (i)~consider different types of workflows based on their resource requirements (e.g., in situ, intra-site, inter-site); (ii)~consider a vertical hierarchical stack starting from the workflow down to computational jobs and individual tasks; (iii)~consider the kind of stakeholder and their needs, etc.

Given the above, an overarching question is how do these different workflow stacks map to different types of workflow systems? Examples of such aspects include resource authentication/authorization and allocation; the granularity of jobs definition/submission: fine-grained (functions) \emph{vs.} coarse-grained (long running jobs); data management (from/to external data stores, within the workflow system, and within the tasks/serverless functions); workflow specification, language, and framework; provenance and error reporting; visualization; and application deployment (native installation \emph{vs.} containerization).

The Global Alliance for Genomics \& Health (GA4GH)~\cite{ga4gh} has targeted the development of API standards (and implementations) that make it easier to ``send the algorithms to the data" on a cloud-based environment. Their workflow stack includes four definitions of standard APIs for interacting with a data repository, tool registry, task execution, and workflow execution services~\cite{ga4gh-cloud}. The Pulsar-Network project~\cite{pulsar-network} provides a distributed, uniform job execution system across European data centers, in which the execution environment is detached from the Galaxy framework. NERSC's SuperFacility API~\cite{enders2020cross} exposes characteristics of the computational facility's resources that could be leveraged by decision making systems. One limitation of these approaches is the need for describing every single component of the stack. When considering heterogeneous platforms (Cloud, HPC, Edge, or the emerging quantum computing environments) the definition of such standards may become overly specialized to a specific platform or service~\cite{gallo2023industrial}. 
The Common Workflow Scheduler~\cite{lehmannHowWorkflowEngines2023} API allows workflow systems to exchange scheduling information with resource managers, making them ``workflow-aware" and capable of scheduling tasks based on workflow dependencies and optimization goals. This approach eliminates the need for workflow systems to implement resource manager-specific logic and simplifies the separation of scheduling responsibilities between workflow systems and resource managers. Currently, this API is available as a plugin for Nextflow and Kubernetes.

\pp{Recommendations}
There is an imminent need for a description of a common workflow stack that might need to be represented by different sets of characterizations depending on the execution environment, however each described using common patterns identified across distinct environments. Once the fundamental workflow stack is defined and understood, the community could focus on the development of standard APIs as the result of a long-term objective.

\subsection{Standardization}
\label{sec:standardization}

Attaining standardization is a foremost requirement to achieve interoperability, portability, and reuse of workflow components and applications. However, full standardization must be accomplished at different levels of the workflow stack. For instance, user-facing standards consider workflow descriptions and workflow input objects, while resource/infrastructure standards target the description of services that can be leveraged by, for example, workflow systems. In addition to standards, shared libraries and designs are also essential for interchanging components and services at horizontal levels of the workflow stack.

One of the challenging aspects of defining standards is the lack of a common terminology that is widely adopted by the community. Unfortunately, the workflows literature suffers from an absence of a common vocabulary that defines terms and describes the components of a workflow. Recent efforts to provide a consolidated view~\cite{hilbrich2022consolidated} or reuse~\cite{suetake2022sapporo} of specification languages for a subset of workflow systems has underlined the discrepancy of term definitions. (Dissimilarities can be even identified at the simple definition of tasks and jobs across workflow systems.) Additionally, it is also necessary to describe the semantics of the components and of the data, as well as the variety of systems on which workflows are executed.

Standardization is also required at the metadata level (user-defined metadata prior to data generation, ad-hoc metadata to annotate a dataset~\cite{skluzacek2021serverless}, and workflow-based metadata)~\cite{blanas2015towards}, which is key for enabling provenance. The Workflow Run RO-Crate working group~\cite{wfrun-ro-crate} is defining profiles for capturing the provenance of an execution of a computational workflow based on different levels of granularity: process (execution of one or more tools that contribute to the same computation), workflow (coordinated execution of the tools driven by a workflow system), and provenance (internal details of each step of the workflow). These profiles are based on the CWLProv profile~\cite{khan2019sharing}, which are organized based on CWL’s workflow mode. Although CWLProv is not a standard such as W3C PROV, it attempts to fill the gaps when translating the latter's definitions to workflow specifications (e.g., traditional standards are hard to adopt into HPC due to their origins in other domains). A remaining open question is how to increase adoption of these standards to other workflow systems, and how to capture specifics of heterogeneous systems without hiding the level of provenance the user cares about?

\pp{Recommendations}
There is a trade-off between sharing technologies \emph{vs.} standards, within constraints that come from computational facilities. As standards are often constraining and/or hard to implement across facilities, one approach could be to focus on promoting standards of giving data from user to workflow systems/operators, i.e. a standard format for describing workflow input objects. Another recommendation is to develop a ``marketplace" for workflow standards that would map them to parts of the workflow stack. The community would then be able to identify relevant standards for their systems and infrastructures, as well as identify gaps in current standards. Finally, there is a pressing need for defining a common vocabulary forcomponents of the workflow stack and even the relationships between them.




\section{AI Workflows}
\label{sec:ai-workflows}

Artificial intelligence (AI) and machine learning (ML) methods are now mainstream in modern science. As computational power increases, including the recent achievement of exascale computing, AI/ML methods can at present, for example, generate highly accurate models that accelerate the rate of scientific discovery of complex problems running at very large scales. As a result, workflows increasingly integrate ML models to guide analysis, couple simulation and data analysis codes, and exploit specialized computing hardware (e.g., GPUs and neuromorphic chips)~\cite{brace2021achieving, EJARQUE2022414}. These workflows inherently couple various types of tasks such as short ML inference, multi-node simulations, and long-running ML model training~\cite{jha2022ai}. They are also often iterative and dynamic, with learning systems deciding in real time how to modify the workflow, e.g., by adding new simulations or changing the workflow all together. In this breakout session, participants focused on understanding the different characteristics of AI workflows, specifically the role of AI to create workflows or when used within workflows.

\subsection{AI Workflows Characteristics}
\label{sec:ai-workflows-characteristics}

In the previous roadmap~\cite{ferreiradasilva2021works}, discussions focused on identifying challenges inherent to workflows in which (most of) their tasks represent AI methods. The main challenges include the fine-grained data management and versioning features, heterogeneity of resources, integration to widely used ML framework, the iterative nature of ML processes, and the support of dynamic branching (Table~\ref{tab:challenges}). 
Although these challenges are still relevant to the current landscape of AI workflows, the community experience with this ever-increasing class of workflows has identified novel characteristics that hinder the efficient execution of these workflow applications at large scales. For instance, managing many small files (e.g., image training sets) may significantly impair the performance of the shared filesystem. Although solutions such as NVMe (nonvolatile memory express) have significantly improved I/O throughput, the data management aspect in the software component still lacks optimized solutions to tackle the large volumes of data produced during workflow execution. Additionally, the volume of metadata generated for provenance may also become intractable. 

In addition to the data management issue above, pseudo-random access to
datasets leads to another major challenge in which optimizing access to
microservices and performing data caching operations becomes a fundamental
scheduling and/or resource provisioning problem. This challenge is aggravated
by the unreliability of models, i.e., training processes often have a
\emph{human-in-the-loop} decision process that may result in a dramatic change
to the shape of the task graph as the workflow execution evolves. Furthermore,
coupling AI-coupled HPC workflows introduces additional challenges arising
from the coupling of AI/ML models to traditional HPC workflows.

\pp{Recommendations}
There is a need for a better understanding of the requirements of AI workflows. Although there exists a diverse set of AI workflow applications that could be leveraged as benchmarks or proxy/mini-apps, the complex specialized deployment (and requirements) of these workflows may prevent their large adoption that would enable comparative studies across computing facilities. Recently, the community has developed workflow benchmarks~\cite{coleman2022pmbs, 8621141, cwl-benchmarks} that help to understand the requirements of traditional workflow applications; however, these benchmarks do not capture most of the characteristics intrinsic to AI workflows. Thus, there is a pressing need for creating a benchmark suite for representative AI workflows.

\subsection{AI Workflows Categories and Motifs}
\label{sec:ai-workflows-categories}

The increasing adoption of AI/ML in modern science has not only enabled the development of novel applications using these methods~\cite{oakes2022building} but also to improvements of systems and problem solving. In the workflows domain, the terminology around `AI workflows' has become overloaded; thus, it is necessary to clearly and consistently define the different (sub)categories that encompass the distinct and unique use of AI/ML methods in the different aspects of workflow applications and systems. One approach to define these workflows into (sub)categories considering the role of AI would be: (i)~\emph{Workflows for AI} -- the workflow is to develop the AI; (ii)~\emph{AI-enabled-workflows}. The latter in turn encapsulates two different sub-classes, viz., the ~\emph{AI-integrated workflows}, where the workflow is about problem-solving with AI; and ~\emph{AI-enhanced-workflow systems}, where the workflow engine is empowered with intelligence using a form of AI.



Within each of the above categories there should be subdivisions that could, for example, consider the dynamic nature of the workflow task graph (conditions, loops) or its adaptive response to events (task prioritization or preemption due to dynamic sampling) that could lead to different design and communication requirements. For the first two categories, the workflows could be classified into applications composed of (i)~\emph{inner}, (ii)~\emph{outer}, or (iii)~\emph{coupled} loops. Examples of inner loops include ML replacement of subroutines, adapting, and parallel training. Outer loops are characterized by reinforcement learning (RL) -- there is a central ``controller" that defines how the different steps (e.g., rollout and training) interact; and active learning -- similar to RL in the sense that there are training, inference, and simulation components with a different class of sampling. Coupled loops can be represented by digital twins, which are a component of the workflow controlled by an AI method.

 A recent characterization used common motifs of AI-HPC workflows~\cite{bethel2023survey} that considers workflows using AI and workflows. These motifs include: (i)~AI models steering ensembles of simulations (e.g., advanced sampling, swarm methods); (ii)~multistage (and typically multiscale and multiphysics) pipelining (e.g., molecule selection, virtual screening); (iii)~inverse design including from observations, or determine causal factors (e.g., molecule or material design given properties such as structure to sequence); (iv)~concurrent duality (e.g., concurrent HPC simulation and AI based digital twins); (v)~distributed models and dynamic data (e.g., distributed AI based reduction/analysis coupled to HPC simulation, diverse models on edge-to-exascale infrastructure); and (vi)~adaptive execution for training (e.g., HPO, NAS, LLM).

\pp{Recommendations}
For each of the above (sub)categories, it is necessary to refine the motifs that would capture the unique requirements of each class; and develop benchmarks based on them that would then help catalyze the development of tailored solutions or example problems. An immediate activity includes spearheading a community effort to formulate a technical/white paper that would provide common terminology and definitions of categories of AI workflows and their unique requirements. As a long-term recommendation, the community needs to determine ways to smooth the integration (along with the deployment) paths across these categories and their associated components.


\section{High Performance Data Management and In Situ Workflows}
\label{sec:insitu}

In situ workflows aim to overcome I/O limitations in HPC by sharing data between simulation, analysis, visualization and orchestration tasks as data are produced. The term ``in situ" has become an umbrella covering approaches well beyond the seminal idea to distribute the load of the traditional post hoc analysis associated with a numerical simulation throughout the execution of that simulation~\cite{childs2020terminology}. Nowadays, in situ workflows facilitate data reduction, annotation, and transformation in different stages (e.g., data acquisition, simulation, analysis, visualization). These workflows can rely on multiple components, execution environments, and data transport methods to bypass the file system while delivering data between components. Additionally, modern in situ workflows exhibit different data production and consumption patterns (i.e., volume, frequency, structure) that need to co-exist in a coordinated and efficient manner. This leads to a need for high-performance management of the input, output, intermediate data, and metadata produced and consumed during the workflow execution.
In this breakout session, participants focused on identifying challenges related to data management for in situ workflows, including how data abstractions may provide fine data management detached from the computing platform, and challenges related to the broader range of in situ workflows today (e.g., streaming and event-based workflows).


In situ workflows have evolved from the traditional simulation and analysis approach. In large-scale science, heavy computations are typically offloaded from the edge where data are produced (e.g., from a large scale scientific instrument such as a particle detector) to HPC infrastructures and cloud resources during burst conditions, leading to complex geographically distributed platforms with diverse performance characteristics. A major challenge in this scenario consists in transferring all this generated data from the edge to the processing facility. An approach to lower the pressure put on network resources is to employ in situ data reduction, an optimization technique in which the size of data is reduced without compromising the quality of the information it carries. This implies that this approach has to control and bound the loss of precision to ensure that end users will still have trust in the data. Approaches leveraging self-descriptive data and metadata could thus be used to define the relevance of pieces of data thanks to a tolerable error bound\cite{mgard, banerjee2022scalable, 7516069} Adaptive compression by regions of interest can also be used to reduce the size of data, using techniques similar to those underlying Adaptive Mesh Refinement: regions where information is more important are compressed less than those where information has little scientific significance. 

The need for high performance data management has significantly increased as modern science applications leverage heterogeneous resources for scaling their computations. The emergence of AI/ML workflows is intertwined with the imminent need for efficiency in situ data management, as these workflows increasingly produce and consume large volumes of data and have new motifs that profoundly differ from traditional workflows (Section~\ref{sec:ai-workflows}). Differences in data formats, acquisition and pre-processing methods, training approaches, and provenance needs further exacerbate the complexity of enabling ML/AI for in situ analysis. Additionally, a key element in modern in situ workflows is the need to integrate them into the edge-to-cloud continuum. In this environment, data exchanges are typically performed through data objects in which the storage stack and data location are hidden from the application. Consequently, individual tasks create and work with objects, not files, and are not concerned with where those objects live. This enables storage-centric optimizations transparent to the end user to improve application performance. A potential solution to address the challenges and requirements of geographically distributed in situ workflows is to develop a data management layer abstraction for providing transparent data operations optimization and intelligent decisions for the workflow application. Ideally, workflow tasks would interact with this data management layer responsible for tracking the location of all these objects and performing intelligent data movements to place the necessary objects physically close to the computations using them. However, providing a single abstraction layer can be challenging when dealing with the requirements of a diverse community (e.g., different types of data, performance requisites, latency, consumption rates, availability needs, etc.). Thus, it is necessary to separate the data management paradigm from the workflow orchestration paradigm and move towards a \emph{data-centric event-driven data plane for workflows}. Note that it is also necessary to detach data from its structure (e.g., file) and associated implementation (e.g., HDF5). Frameworks such as ADIOS2~\cite{GODOY2020100561} and Maestro~\cite{maestro}, combined with self-descriptive data (metadata) capabilities and the publish/subscribe model, can be used to expose the data interaction mechanisms available to applications when building workflows.


Emerging in situ workflows involve streaming and event-based elements. Event-driven workflows have a strong need for tightly-coupled and high performing components that integrate HPC and streaming (especially for urgent computing)~\cite{9307968}. In these workflows, in situ computations are initiated based on events (e.g., changes to a variable or data, or access in a data repository). This class of workflows has been largely adapted to Cloud computing environments based on a service-oriented approach that uses streams of data for composing the workflows. In the HPC environment, streaming workflows are typically limited by policies that restrict connectivity to external data sources or long-running services. One approach to address these limitations is to couple a container-based platform (e.g., NERSC's Spin, OLCF's Slate) to HPC systems, in which services running on these platforms would act as a bridge between the outside world and the HPC ecosystem. Another limitation for enabling urgent computing on HPC is the need for specialized queues that can swiftly allocate resources for these types of jobs.

\pp{Recommendations}
There is a need for determining what are the requirements for upcoming workflow applications, especially those integrating data or moving data across facilities. Dataflow and I/O contention benchmarks are needed to assist in the quantification of performance for workflow and data management solutions in upcoming scenarios. Current efforts~\cite{coleman2022pmbs, wci-benchmark} mostly target traditional workflow motifs, and must be expanded to cover new applications (e.g., AI/ML workflows) and infrastructures (e.g., edge-to-cloud workflows). There is also a need for defining community-driven schemas to describe data and metadata, and to further provide an integrated view of the different views of the data associated to a workflow execution.

There is also a need to define a service-oriented composability approach for enabling urgent computing workflows in the HPC ecosystem. To explore this, the community should conduct a comprehensive study to understand the specific needs for this class of applications for HPC, derive lessons learned from the service-oriented Cloud computing solutions, and engage with HPC computing facility operators to design a potential solution that would satisfy both the application's performance requirements as well as guarantee that facilities' security policies are still enforced.


\section{HPC and Quantum Workflows}
\label{sec:hpc-quantum}

Most current workflow tools can operate over classical HPC, providing automated orchestration of tasks and data management. With the emergence of quantum computing (QC) there is a need to bridge these two computing models to improve the efficiency and potential impact of applications that could leverage the capabilities of each model. The need for a hybrid QC-HPC/Cloud approach is motivated for modern applications. In the quantum chemistry domain, for instance,researchers seek to understand and harness the quantum properties of atoms and systems around us. QC has also been increasingly used for ML processing and optimization -- conversely, ML has been leveraged for the calibration of quantum processors. 

In the current state of quantum computing, a.k.a. noisy intermediate-scale quantum (NISQ) era, there is a limited number of noisy physical qubits, thus it is paramount that quantum algorithms be efficient. Recently, variational quantum algorithms (VQAs)~\cite{cerezo2021variational} have shown modest success in ML and optimization tasks. The approach requires a constant exchange of data between classical and quantum devices, which leads to a model in which quantum devices act as accelerators for classical computing. As a result, managing these sets of distributed, heterogeneous devices motivates the need for workflow solutions that can provide a common abstraction layer to interface with the multitude of specialized APIs and components provided by each platform. 

Due to the novelty of the topic, this breakout session was structured with a series of short talks that underlined different aspects of quantum computing in workflows research, and identified a list of challenges associated to each of these aspects. Table~\ref{tab:quantum-topics} shows the list of topics. All presentation videos can be found at the WCI YouTube channel (\url{https://www.youtube.com/playlist?list=PLAtmuqHExRvO-l-mACmlK_4mfmG6yLnpN}).

\begin{table}[!ht]
    \centering
    \small
    \setlength{\tabcolsep}{12pt}
    \begin{tabular}{p{7.5cm}p{7cm}}
        \toprule
        \textbf{Topic} & \textbf{Presenter} \\
        \midrule
        \rowcolor[HTML]{F2F2F2}
        Tensor Networks for Quantum Simulation & Artur Garcia Saez (BSC) \\
        Advances in Hybrid Quantum-Classical High-Performance Computing & Stefan Andersson (ParTec AG), Mathias Pütz (ParTec AG) \\
        \rowcolor[HTML]{F2F2F2}
        Workflow Scheduling Using Quantum Devices & Justyna Zawalska (CYFRONET AGH) \\
        Mechanisms for Enhancing Reliability/Recovery and Performance of Future Workloads & Ryan E. Grant (Queen’s University) \\
        \rowcolor[HTML]{F2F2F2}
        HPCQC System Workflows & Martin Ruefenacht (LRZ) \\
        Multi-core Quantum Computing & Aharon Brodutch (Entangled Networks) \\
        \bottomrule
    \end{tabular}
    \caption{HPC and quantum workflows breakout topics.}
    \label{tab:quantum-topics}
\end{table}

\subsection{State-of-the-art and Challenges}

Recent efforts target the definition of languages for modeling computations between classical and quantum computing~\cite{weder2020integrating}. The open-source Covalent framework~\cite{covalent} provides mechanisms to manage experiments (expressed as workflows) and facilitate access to quantum devices, bridging them to HPC and cloud platforms. RosneT is a library for distributed, out-of-core block tensor algebra~\cite{sanchez2021rosnet} built on top of the PyCOMPSs~\cite{tejedor2017pycompss} programming model to transform tensor operations into a collection of tasks to be executed on HPC resources. QC has also been leveraged to solve well-known scheduling problems for traditional workflow applications~\cite{wf-quantum-scheduling}.

Despite the above efforts, there are still several challenges that need to be addressed. For instance, (i)~the reproducibility aspect in quantum computing requires domain expertise to understand the circuits and qubit topologies; (ii)~the scheduling timescales significantly differ between classical and quantum computing -- the latter is performed in the order of microseconds; (iii)~there is a limited number of available resources, thus waiting time in queues can attain hours and costs to perform substantial computations may become excessive; (iv)~programming each quantum device requires expert knowledge for each vendor API; (v)~there are no standard representations for intermediate representations (IRs) for quantum programs, which are described as circuits; (vi)~although some approaches bridge quantum devices to HPC and cloud systems, there is no tight integration between them (e.g., as between GPUs and CPUs).

\pp{Recommendations}
The community is actively working in multiple objectives to make progress in the different topics of research. The foremost research directions include (i)~the optimization of classical simulators based on tensor networks in classical HPC systems; (ii)~the need to develop quantum systems with a component-based architecture to facilitate the integration with HPC and cloud environments; (iii)~moving the hardware-aware software from the Quantum to the HPC system; (iv)~reaching a consensus between QC as accelerators for classical computing (e.g., QPUs as accelerators to combine traditional HPC-QC workflows) or \emph{vice versa}; (v)~predicting optimizability of workflow decisions (e.g., how long will it take to optimize a QC workflow? Should a circuit be optimized?).


\section{FAIR Computational Workflows}
\label{sec:fair}

The original FAIR principles~\cite{wilkinson2016fair} laid a foundation for sharing and publishing data assets, emphasizing machine accessibility in that data and all other assets should be: (i)~Findable -- user of persistent identifies, cataloguing and indexing of data; (ii)~Interoperable -- machine processable metadata using standards; (iii)~Accessible -- clear access protocols to access (meta)data; (iv)~Reusable -- metadata standards, machine accessible usage license and provenance. When considering FAIR computational workflows, both data and software aspects need to be considered. FAIR data principles can be applied to, for example, workflow descriptions and specifications (typically via a domain specific language or API) that can be associated to metadata with Digital Object Identifiers (DOIs) or objects such as test data and parameter files. Workflow software objects, on the other hand, bring additional challenges such as reproducibility, usability, quality, maturity, etc. The recently published FAIR for research software principles~\cite{hong2022fair} can then be applied to both workflows and workflow management systems. In this breakout session, participants focused on the software aspect of workflows and discussed best practices for research software registries and repositories, as well as approaches to build FAIR into workflow systems.

\subsection{FAIR Workflow Repositories and Registries}

The current state of the art for finding and accessing workflows consists of community repositories (e.g., nf-core~\cite{ewels2020nf}, snakemake workflow catalog~\cite{swc}), community platforms (e.g., nextflow tower~\cite{nft}), data repositories (e.g., Zenodo~\cite{zenodo}, Dataverse~\cite{trisovic2020advancing}), registries (e.g., WorkflowHub~\cite{goble2021implementing}, Dockstore~\cite{dockstore}), and metadata frameworks (e.g., CWL~\cite{crusoe2022methods}, Workflow RO-Crate~\cite{workflow-ro-crate}, Bioschemas profiles~\cite{bioschema}). Registries and repositories provide curation and best practices for recording workflows, while metadata frameworks capture workflows description and their associated metadata (e.g., via canonical descriptions).

The notion that workflows can encompass both data and software brings several challenges when building FAIR workflow repositories and registries. In addition to capturing the workflow descriptions and (reference to) their input data, it is also necessary to capture metadata, containers, execution information, configurations, etc. The current approach adopted by the community is to separate the workflow definition from its execution -- e.g., WorkflowHub records the workflow description and refers to Workflow-RO-Crates  objects for execution information; Workflow-Run-RO-Crates collect the actual execution provenance. Although this approach has demonstrated success for traditional DAG- and cloud-based workflows, capturing workflow executions in the edge-to-HPC computing continuum is still an open question -- mostly due to their intrinsic configurations, specialized architectures, and unique scientific instruments. The dynamic nature of emerging AI/ML workflows poses an additional challenge to capture executions that match the canonical description of the workflow.  

An additional feature for registries would be to provide mechanisms to characterize workflows in terms of structures so that a user could explore similar solutions to their problem. Recent approaches on extracting metadata from software and research data repositories~\cite{filgueira2022inspect4py, kelley2021framework, skluzacek2021serverless} or automated extraction of workflow patterns~\cite{coleman2021escience} could be leveraged to measure similarities across workflows. Past work~\cite{starlinger2014similarity} should be revisited. For instance, workflows could be compared by tags or labels, by the workflow structure (i.e., task graph), or by a stored relationship (e.g., sub-workflow, forked variant, etc.). There is also a need for defining persistent identifiers (PIDs) for workflows. DOIs are appropriate for workflows that are ``completed and published", i.e., as a snapshot of the workflow -- similar to the publication process of research articles. There is a need for defining persistent identifiers (PIDs) for workflows that capture their complex nature.

\pp{Recommendations}
Given the above, an imminent need is to define standards for capturing the metadata for workflows. This is critical for enabling automatic FAIR (unit/end-to-end) testing of workflows to support reuse and composability. There is also a need to expand the first set of recommendations for research software registries and repositories~\cite{garijo2022nine} for scientific workflows, especially attempting to address the need for workflow PIDs. A long-term activity is to consider new paradigms for workflow descriptions that could capture the emerging class of dynamic workflows.

\subsection{Building FAIR into Workflow Management Systems}

In addition to providing repositories and registries to store workflows and their associated metadata/information, it is crucial to empower workflow systems with capabilities to enable support to FAIR data and software throughout the workflow execution. To attain this objective, the community argues that two key features are needed: \emph{standards} and \emph{metadata}. Specifically, there is a need for a standard for expressing the inputs of the workflow and how to set them. Unfortunately, this issue is not novel and has not received much attention from the community. In fact, this issue is more a people problem than a technical problem, as is frequently also the case when attempts to implement FAIR practices fall short~\cite{the-f-paper}. Ideally, in terms of FAIRness, documentation is the most important asset for a workflow. It is very difficult to maintain software, but documentation lives on. On the description of the workflow itself, there are system agnostic languages (e.g., CWL and OpenWDL~\cite{openwdl}) that have been adopted by several workflow systems to foster workflow portability, though these may be limited in the workflows that they can describe. The goal is to provide a mechanism to describe workflows so that they can be compared. Portability at the system level remains an open question, especially in edge and HPC environments.

The availability of metadata is another key feature necessary to enable FAIR within workflow systems. To this end, FAIR registration needs to be automatic and cannot introduce overhead in recording provenance information (i.e., the facts that link the inputs and outputs of the workflow) when running the workflow -- recording of provenance information needs to be transparent and scalable~\cite{sirvent2022automatic}. Thus, FAIR data and FAIR workflows are intertwined. A solution would be to leverage Workflow-Run RO-Crate profiles, where the workflow system would export these profiles that would capture the provenance of an execution of a computational workflow. As a result, a coupling between the workflow and its associated data and metadata would be properly documented.

\pp{Recommendations}
An immediate action is to continue the efforts to define methods for fostering portability across workflow systems, instead of defining the ``standard" language for expressing workflows. Another key direction is to invest into people and communities rather than specific tools. The community strongly believes that the development of people's and communities' skills will outlast workflow systems, thus the incentive for FAIRness will be built intrinsically into these systems. Last, there is a need to assess the limitation of FAIR for different kinds of workflow types such as streaming and IoT workflows, or HPC workflows that are monolithic and tied to a particular architecture.


\section{Workflows for Continuum and Cross-Facility Computing}
\label{sec:cross-facility}

Continuum and cross-facilities workflows are becoming more prevalent in the computational sciences. As these workflow paradigms have recently emerged, their definitions are evolving. The current understanding is that continuum workflows represent analysis pipelines that require continuous access to computing (e.g., urgent computing). Examples of workflows that fall under this category include in situ workflows, components that talk to each other within the same system (Section~\ref{sec:insitu}); geographical computing, i.e., edge-to-cloud-to-HPC computing; and high throughput computing, e.g., long-running computing campaigns over long periods of time. Cross-facility workflows represent analysis pipelines that hit more than a single site, which may include an experiment and a computing site, multiple experiment sites, or multiple computing facilities~\cite{antypas2021enabling} -- a site can be defined as a local compute, HPC, cloud, edge, campus cluster, and sensors at the edge. Cross-facility workflows can be seen as a solution for the needs of a continuum workflow (e.g., commercial cloud providers can be considered effectively cross-facility) offering resiliency for the computing needs of real time workflows. Another view is to consider continuum and cross-facility computing by the data plane, i.e., workflow systems should be able to tolerate different representations and underlying storage systems. In this breakout session, participants focused on identifying challenges to enable these categories of workflows and potential actions for the community.

\subsection{Characteristics and Challenges}

Data management is a major challenge in modern science, which involves operating data streams (e.g., from an instrument to a computing resource), data staging (e.g., processing fragments of data from different locations), or data movement (e.g., wide area transfers). The need for running workflows on multiple computing sites has increased as, for example, nowadays instruments can generate an ever-growing volume of data (in the order of TB/s)~\cite{mcclure2020}. These workflows are characterized by their unique requirements to access and operate over computing resources regardless of the time or location sensitivity. For instance, existing computing facilities provide no mechanisms for scheduling and/or provisioning I/O capacity. Enabling a data-centric approach (i.e., doing computation on data quanta) may make the problem more tractable. To this end, it is necessary that tools to measure system contention/resource utilization and capture I/O and data movement performance are made available for measuring this end-to-end performance at multiple levels: computing system, across the network (local and wide area), and across the multiple compute and user facilities (e.g., node memory at one site to a file system at another site). Current systems perform matchmaking of task requirements (CPU/GPU/RAM) across multiple HPC platforms (e.g., RADICAL-Cybertools~\cite{balasubramanian2019radical}, HTCondor~\cite{thain2005distributed}, etc.), however I/O requirements are neglected due to the lack of control over the I/O performance of an HPC machine at a given point in time.

A precondition to attain the envisaged level of fine-grained workflow orchestration described above is to define the requirements of a workflow task. Specifically, what metadata does the workflow task need to have to request resources (computing, storage, network) from the appropriate site? To answer this question, it is necessary to convey that tasks are less portable than commonly assumed (including containers). The process for labelling every single resource a task needs is still an open question, and requirements may significantly diverge considering the targeted computing environment (e.g., compiler optimization for a determined architecture, data location and network and I/O bandwidths, policies, etc.). 

In addition to the above challenges, cross-facility workflows also face challenges associated to autonomous administration domains, i.e., different policies and security models at different sites (authorization, access control), different software stacks, etc. Approaches for providing resource information in a machine-readable format (e.g., NERSC's superfacility project~\cite{enders2020cross}) or enabling federated identification (e.g., DOE's OneID~\cite{oneid}) are a first step for enabling distributed workflow orchestration and intelligent decisions at the workflow management system level. A collaboration between the Computational Science Initiative (CSI), the Center for Functional Nanomaterials (CFN), and the National Synchrotron Light Source (NSLS-II) attempt to facilitate using larger institutional compute resources to support the beamlines. The idea is to go from giving users access to beamlines and sending them home with their data, to giving users access to experiments and sending them home with answers to their science questions. Multi-site workflows are a mainstream component to this effort.

A different approach is letting users embrace and exploit heterogeneity from the design phase through \textit{hybrid workflows}~\cite{streamflow:21}. Traditional workflow models, which describe steps and data dependencies, can be flanked by a topology of execution locations, encoding execution environments, available resources, and communication channels. Users can then explicitly map the different steps of the workflow onto different locations of the topology in an N-to-M relationship. Hybrid workflows are heterogeneous by design and do not require a homogeneous representation of locations' capabilities. As a result, site-specific plugins can be used to handle orchestration aspects such as authentication, scheduling, data transfers, and task offloading.
Hybrid workflows have already successfully orchestrated cloud-HPC and cross-HPC distributed workflows using batched~\cite{streamflow:21} and interactive~\cite{jupyter-workflow:22} execution paradigms. The main drawback is that the additional complexity introduced in the design phase steepens the learning curve for domain experts with little knowledge of distributed execution environments.

\pp{Recommendations}
Continuum and cross-facility workflows are becoming mainstream in modern science. These new classes of workflows present, in addition to the typical challenges inherent to computing on a single site, new challenges that require coordination and cooperation among computing and experimental facilities. An immediate action is to enable tracking the metadata associated with tasks in workflow across different sites. To this end, it is necessary to define a standard/specification for task descriptions, including a way to represent the overall I/O requirements of an entire workflow. There is also a need for a community effort to define what a computing site needs to provide to enable a workflow orchestrator to make intelligent decisions about where to place tasks in a workflow. Another recommendation is to factor in the need for cross-facility computing from the conceptualization/design phase of computer systems and experimental facilities -- typically, personnel involved in the design of experiment facilities are not necessarily computing experts.

\newpage
\cleardoublepage\phantomsection\addcontentsline{toc}{section}{References}
\bibliographystyle{IEEEtranDOI}
\bibliography{references}

\newpage
\cleardoublepage\phantomsection\addcontentsline{toc}{section}{Appendix A: Participants and Contributors}
\section*{Appendix A: Participants and Contributors}
\label{appx:contributors}

\fontsize{10}{11}\selectfont
\begin{longtable}[!h]{llp{10.5cm}}
\toprule
\textbf{First Name} & \textbf{Last Name} & \textbf{Affiliation} \\
\midrule
\rowcolor[HTML]{F2F2F2} 
Ilkay         & Altintas           & UC San Diego \\
Stefan        & Andersson          & ParTec AG \\
\rowcolor[HTML]{F2F2F2} 
William       & Arndt              & National Energy Research Scientific Computing Center \\
Juan          & Aznar              & University of Innsbruck \\
\rowcolor[HTML]{F2F2F2} 
Jonathan      & Bader              & TU Berlin \\
Rosa M        & Badia              & Barcelona Supercomputing Center \\
\rowcolor[HTML]{F2F2F2} 
Venkat        & Bala               & Agnostiq \\
Bartosz       & Balis              & AGH University of Science and Technology\\
\rowcolor[HTML]{F2F2F2} 
Debbie        & Bard               & National Energy Research Scientific Computing Center \\
Chris         & Blanton            & NOAA-GFDL \\
\rowcolor[HTML]{F2F2F2} 
Timo          & Bremer             & Lawrence Livermore National Laboratory \\
Aharon        & Brodutch           & Entangled Networks \\
\rowcolor[HTML]{F2F2F2} 
Ben           & Brown              & DOE ASCR \\
Paul          & Brunk              & University of  Georgia \\
\rowcolor[HTML]{F2F2F2} 
Ian           & Buckley            & Agnostiq \\
Silvina       & Caino-Lores        & University of Tennessee \\
\rowcolor[HTML]{F2F2F2} 
Henri         & Casanova           & University of Hawaii at Manoa \\
Alba          & Cervera-Lierta     & Barcelona Supercomputing Center \\
\rowcolor[HTML]{F2F2F2} 
Kyle          & Chard              & University of Chicago \\
Justin        & Chigu              & Egypt Japan University of Science and Technology \\
\rowcolor[HTML]{F2F2F2} 
Taina         & Coleman            & University of Southern California \\
Nick          & Collier            & Argonne National Laboratory \\
\rowcolor[HTML]{F2F2F2} 
Iacopo        & Colonnelli         & Universit\`{a} degli Studi di Torino \\
Frederik      & Coppens            & ELIXIR \\
\rowcolor[HTML]{F2F2F2} 
Michael       & Crusoe             & CWL Project / VU Amsterdam / ELIXIR NL \& DE \\
Will          & Cunningham         & Agnostiq \\
\rowcolor[HTML]{F2F2F2} 
Bruno         & de Paula Kinoshita & Barcelona Supercomputing Center \\
Paolo         & Di Tommaso         & Seqera Labs \\
\rowcolor[HTML]{F2F2F2} 
Charles       & Doutriaux          & Lawrence Livermore National Lab \\
Matthew       & Downton            & NCI Australia \\
\rowcolor[HTML]{F2F2F2} 
Wael          & Elwasif            & Oak Ridge National Laboratory \\
Bjoern        & Enders             & NERSC \\
\rowcolor[HTML]{F2F2F2} 
Chris         & Erdmann            & Michael J. Fox Foundation \\
Thomas        & Fahringer          & University of Innsbruck \\
\rowcolor[HTML]{F2F2F2} 
Rafael        & Ferreira da Silva  & Oak Ridge National Laboratory \\
Ludmilla      & Figueiredo         & German Centre for Integrative Biodiversity Research \\
\rowcolor[HTML]{F2F2F2} 
Rosa          & Filgueira          & University of St. Andrews \\
Martin        & Foltin             & HP Labs \\
\rowcolor[HTML]{F2F2F2} 
Anne          & Fouilloux          & Simula Research Laboratory \\
Luiz          & Gadelha            & University of Jena \\
\rowcolor[HTML]{F2F2F2} 
Andy          & Gallo              & GE Research \\
Artur         & Garcia             & Barcelona Supercomputing Center \\
\rowcolor[HTML]{F2F2F2} 
Daniel        & Garijo             & Universidad Politécnica de Madrid \\
Roman         & Gerlach            & Friedrich-Schiller-University Jena, Germany \\
\rowcolor[HTML]{F2F2F2} 
Carole        & Goble              & The University of Manchester \\
Ryan          & Grant              & Queen's University \\
\rowcolor[HTML]{F2F2F2} 
Samuel        & Grayson            & University of Illinois at Urbana-Champaign \\
Patricia      & Grubel             & Los Alamos National Laboratory \\
\rowcolor[HTML]{F2F2F2} 
Johan         & Gustafsson         & Australian BioCommons, University of Melbourne \\
Valerie       & Hayot              & University of Chicago \\
\rowcolor[HTML]{F2F2F2} 
Oscar         & Hernandez          & ORNL \\
Marcus        & Hilbrich           & Humboldt-Universität zu Berlin \\
\rowcolor[HTML]{F2F2F2} 
Shantenu      & Jha                & Brookhaven National Laboratory / Rutgers University \\
Annmary       & Justine            & Hewlett Packard Enterprise \\
\rowcolor[HTML]{F2F2F2} 
Daniel S.     & Katz               & University of Illinois at Urbana-Champaign \\
Ian           & Laflotte           & SAIC / NOAA/ GFDL \\
\rowcolor[HTML]{F2F2F2} 
Daniel        & Laney              & Lawrence Livermore National Laboratory \\
Fabian        & Lehmann            & Humboldt-Universität zu Berlin \\
\rowcolor[HTML]{F2F2F2} 
Andre         & Luckow             & Ludwig-Maximilian University Munich \\
Jakob         & Luettgau           & University of Tennessee \\
\rowcolor[HTML]{F2F2F2} 
Ketan         & Maheshwari         & Oak Ridge National Laboratory \\
Motohiko      & Matsuda            & RIKEN \\
\rowcolor[HTML]{F2F2F2} 
Doriana       & Medic              & University of Turin \\
Pete          & Mendygral          & HPE \\
\rowcolor[HTML]{F2F2F2} 
Marek         & Michalewicz        & Sano Centre for Personalized Computational Medicine \\
Jorji         & Nonaka             & RIKEN R-CCS \\
\rowcolor[HTML]{F2F2F2} 
Manish        & Parashar           & University of Utah \\
Maciej        & Pawlik             & University of Science and Technology AGH \\
\rowcolor[HTML]{F2F2F2} 
Loïc          & Pottier            & Lawrence Livermore National Laboratory \\
Line          & Pouchard           & Brookhaven National Lab \\
\rowcolor[HTML]{F2F2F2} 
Mathias       & Pütz               & ParTec AG \\
Santosh Kumar & Radha              & Agnostiq Inc. \\
\rowcolor[HTML]{F2F2F2} 
Lavanya       & Ramakrishnan       & LBNL \\
Sashko        & Ristov             & University of Innsbruck \\
\rowcolor[HTML]{F2F2F2} 
Paul          & Romano             & Argonne National Laboratory \\
Kelly         & Rosa Braghetto     & University of São Paulo \\
\rowcolor[HTML]{F2F2F2} 
Daniel        & Rosendo            & Inria \\
Martin        & Ruefenacht         & Leibniz Supercomputing Centre \\
\rowcolor[HTML]{F2F2F2} 
Katarzyna     & Rycerz             & II \& Cyfronet AGH \\
Nishant       & Saurabh            & Utrecht University \\
\rowcolor[HTML]{F2F2F2} 
Volodymyr     & Savchenko          & EPFL \\
Martin        & Schulz             & TU-Munich and LRZ \\
\rowcolor[HTML]{F2F2F2} 
Christine     & Simpson            & ALCF \\
Raül          & Sirvent            & Barcelona Supercomputing Center \\
\rowcolor[HTML]{F2F2F2} 
Tyler         & Skluzacek          & Oak Ridge National Laboratory \\
Stian         & Soiland-Reyes      & The University of Manchester \\
\rowcolor[HTML]{F2F2F2} 
Renan         & Souza              & Oak Ridge National Lab \\
Sreenivas Rangan & Sukumar         & HPE \\
\rowcolor[HTML]{F2F2F2} 
Ziheng        & Sun                & George Mason University \\
Alan          & Sussman            & University of Maryland \\
\rowcolor[HTML]{F2F2F2} 
Frederic      & Suter              & ORNL \\
Douglas       & Thain              & Notre Dame University \\
\rowcolor[HTML]{F2F2F2} 
Mikhail       & Titov              & Brookhaven National Laboratory \\
Benjamin      & Tovar              & Univeristy of Notre Dame \\
\rowcolor[HTML]{F2F2F2} 
Aalap         & Tripathy           & Hewlett Packard Enterprise \\
Matteo        & Turilli            & BNL; Rutgers University \\
\rowcolor[HTML]{F2F2F2} 
Bartosz       & Tużnik             & ICM, University of Warsaw \\
Nick          & Tyler              & NERSC \\
\rowcolor[HTML]{F2F2F2} 
Thomas        & Uram               & Argonne National Laboratory \\
Hubertus      & van Dam            & Brookhaven National Laboratory \\
\rowcolor[HTML]{F2F2F2} 
Aurelio       & Vivas              & Universidad de los Andes \\
Logan         & Ward               & Argonne National Laboratory \\
\rowcolor[HTML]{F2F2F2} 
Patrick       & Widener            & Oak Ridge National Laboratory \\
Sean          & Wilkinson          & Oak Ridge National Laboratory \\
\rowcolor[HTML]{F2F2F2} 
Justyna       & Zawalska           & AGH University of Science and Technology, ACC Cyfronet \\
Mahnoor       & Zulfiqar           & Friedrich Schiller University \\ 
\bottomrule                                       
\end{longtable}

\fontsize{11}{11}\selectfont

\newpage
\cleardoublepage\phantomsection\addcontentsline{toc}{section}{Appendix B: Agenda}
\section*{Appendix B: Agenda}
\label{appx:agenda}

\vspace{10pt}
\noindent
\textbf{Day 1 -- November 29, 2022}
\fontsize{10}{11}\selectfont
\begin{longtable}{lp{12cm}}
    \toprule
    \textbf{Time} & \textbf{Topic} \\
    \midrule
    11:00-11:05am EST & \makecell[l]{\textbf{Welcome and Introductions}\\
    \emph{\small Rafael Ferreira da Silva (Oak Ridge National Laboratory)}} \\
    
    \rowcolor[HTML]{EEEEEE}
    11:05-11:10am EST & \makecell[l]{\textbf{The Workflows Community Initiative overview}} \\

    11:10-11:20am EST & \makecell[l]{\textbf{Remembering the 2021 Roadmap}\\
    \emph{\small Rafael Ferreira da Silva (Oak Ridge National Laboratory)}} \\

    \rowcolor[HTML]{EEEEEE}
    11:20-11:50am EST & \textbf{Lightning talks}
        \begin{compactitem}
            \item Topic 1: Specifications, standards, and APIs
            \item[] \emph{Daniel Laney (LLNL), Kyle Chard (UChicago)}
            \item Topic 2: AI workflows 
            \item[] \emph{Shantenu Jha (BNL), Timo Bremer (LLNL)}
            \item Topic 3: High performance data management and in situ workflows 
            \item[] \emph{Silvina Caino-Lores (UTK), Fr\'ed\'eric Suter (ORNL)}
            \vspace{-10pt}
        \end{compactitem} \\
        
    11:50am-noon EST & \emph{10min Break} \\
    
    \rowcolor[HTML]{EEEEEE}
    noon-1:30pm EST & \textbf{Breakout sessions}
    \begin{compactitem}
            \item Topic 1: Specifications, standards, and APIs
            \item Topic 2: AI workflows 
            \item Topic 3: High performance data management and in situ workflows 
            \vspace{-10pt}
        \end{compactitem} \\
    1:30-1:45pm EST & \emph{15min Break} \\

    \rowcolor[HTML]{EEEEEE}
    1:45-2:00pm EST & \textbf{Reports from breakout sessions} \\
    \bottomrule
\end{longtable}

\vspace{20pt}
\fontsize{11}{11}\selectfont
\noindent
\textbf{Day 2 -- November 30, 2022}
\fontsize{10}{11}\selectfont
\begin{longtable}{lp{12cm}}
    \toprule
    \textbf{Time} & \textbf{Topic} \\
    \midrule
    11:00-11:10am EST & \makecell[l]{\textbf{Welcome and Introductions}\\
    \emph{\small Rafael Ferreira da Silva (Oak Ridge National Laboratory)}} \\
    
    \rowcolor[HTML]{EEEEEE}
    11:10-11:45am EST & \textbf{Lightning talks}
        \begin{compactitem}
            \item Topic 4: HPC and Quantum workflows 
            \item[] \emph{Rosa Badia (BSC), Ian Buckley (Agnostiq), Venkat Bala (Agnostiq)}
            \item Topic 5: FAIR workflows
            \item[] \emph{Carole Goble (UManchester), Daniel S. Katz (UIUC)}
            \item Topic 6: Workflows for continuum and cross-facility computing 
            \item[] \emph{Debbie Bard (NERSC), Tom Uram (ANL), Nick Tyler (NERSC)}
            \vspace{-10pt}
        \end{compactitem} \\
        
    11:45am-noon EST & \emph{15min Break} \\

    \rowcolor[HTML]{EEEEEE}
    noon-12:15pm EST & \makecell[l]{\textbf{Invited talk:} A Translational Perspective on End-to-End Workflows\\
    \emph{\small Manish Parashar (Utah)}} \\
    
    12:15-1:35pm EST & \textbf{Breakout sessions}
    \begin{compactitem}
            \item Topic 4: HPC and Quantum workflows 
            \begin{compactitem}
                \item Tensor Networks for Quantum Simulation 
                \item[] \emph{Artur Garcia Saez (BSC)}
                \item Advances in Hybrid Quantum-Classical High-Performance Computing 
                \item[] \emph{Stefan Andersson (ParTec AG), Mathias Pütz (ParTec AG)}
                \item Workflow Scheduling Using Quantum Devices 
                \item[] \emph{Justyna Zawalska (Academic Computer Centre CYFRONET AGH)}
                \item Mechanisms for Enhancing Reliability/Recovery and Performance of Future Workloads 
                \item[] \emph{Ryan E. Grant (Queen's University)}
                \item HPCQC System Workflows 
                \item[] \emph{Martin Ruefenacht (Leibniz Supercomputing Centre)}
                \item Multi-Core Quantum Computing 
                \item[] \emph{Aharon Brodutch (Entangled Networks)}
            \end{compactitem}
            \item Topic 5: FAIR workflows
            \item Topic 6: Workflows for continuum and cross-facility computing 
            \vspace{-10pt}
        \end{compactitem} \\

    \rowcolor[HTML]{EEEEEE}
    1:30-1:45pm EST & \emph{15min Break} \\

    1:45-2:00pm EST & \textbf{Reports from breakout sessions} \\
    \bottomrule
\end{longtable}

\end{document}